\documentclass[conference]{IEEEtran}
\usepackage{spconf,amsmath,graphicx}

\usepackage{hyperref}

\title{Emotion-Coherent Speech Data Augmentation and Self-Supervised Contrastive Style Training for Enhancing Kids's Story Speech Synthesis}
\name{Raymond Chung}
\address{
    Logistics and Supply Chain MultiTech R\&D Centre, Pok Fu Lam, Hong Kong \\    
    \ \tt rchung@lscm.hk
}

\begin{document}
%

\IEEEoverridecommandlockouts
\IEEEpubid{\makebox[\columnwidth]{978-1-5386-5541-2/18/\$31.00~\copyright2018 IEEE \hfill}
\hspace{\columnsep}\makebox[\columnwidth]{ }}
\maketitle
\IEEEpubidadjcol

\begin{center}
    \vspace{-1em}
    \small
    © 2024 IEEE. This is the author’s version of the work. 
    The final published paper is available at 
    \href{https://doi.org/10.1109/SLT61566.2024.10832287}{DOI: 10.1109/SLT61566.2024.10832287}.
\end{center}

\begin{abstract}
Expressive speech synthesis requires vibrant prosody and well-timed pauses. We propose an effective strategy to augment a small dataset to train an expressive end-to-end Text-to-Speech model. We merge audios of emotionally congruent text using a text emotion recognizer, creating augmented expressive speech data. By training with two-sentence audio, our model learns natural breaks between lines. We further apply self-supervised contrastive training to improve the speaking style embedding extraction from speech. During inference, our model produces multi-sentence speech in one step, guided by the text-predicted speaking style. Evaluations showcase the effectiveness of our proposed approach when compared to a baseline model trained with consecutive two-sentence audio. Our synthesized speeches give a closer inter-sentence pause distribution to the ground truth speech. Subjective evaluations reveal our synthesized speech scored higher in naturalness and style suitability than the baseline.
\end{abstract}

\section{Introduction}
\label{sec:intro}

Producing an audiobook typically involves a voice talent reading the entire book in a professional studio, often with a partner to monitor and assist during the recording process. Machine learning-based speech synthesis holds significant potential to revolutionize this resource-intensive audiobook creation process.




Recent advancements in text-to-speech (TTS) models, such as Tacotron 2 ~\cite{taco2} and FastSpeech ~\cite{ ren2020fastspeech}, have resulted in highly natural-sounding speech for short sentences, typically lasting around 10 seconds. This brevity can be traced back to several factors: the training datasets, like LJS~\cite{ljspeech17} and LibriTTS~\cite{zen2019libritts}, are segmented into short lengths, and these short lengths allow for larger patch sizes, enabling more efficient model training. Additionally, auto-regressive models struggle with generating longer outputs. To overcome this limitation, He et al. \cite{he2019robust} proposed a stepwise monotonic attention, while Battenberg et al.  \cite{battenberg2020location} introduced a dynamic convolution attention. Both of these approaches have successfully enabled the generation of longer speech outputs, paving the way for long-form TTS applications. 

The 2017 Blizzard Machine Learning Challenge\cite{sawada2017blizzard} presented a unique opportunity by providing a small dataset of 6.5 hours of children's audiobooks narrated by a professional British female speaker. Out of the seven registered teams, only three submitted their statistical speech synthesis systems, as the task design posed challenges for implementing end-to-end approaches. A released version of this dataset, known as "NITECH," is organized by dividing the audiobook content by page. Despite the availability of this well-structured and highly expressive dataset, it has received limited research attention, likely due to the small size of the training data. To address this data constraint, we pre-trained the proposed TTS model using the LibriTTS dataset~\cite{zen2019libritts}.


In this paper, we propose a TTS model to generate one audio for each page of the kid storybooks, in which each page usually consists of multiple sentences. While a TTS model could read text naturally, we focus on improving the training of the expressiveness module of a TTS model. We introduce a unique data augmentation method and self-supervised contrastive training to refine this module’s training, which can be seamlessly incorporated into existing research. Our innovations include:


\begin{enumerate}
  \item We propose a distinctive data augmentation approach that leverages a fine-tuned emotion text classifier, originating from a self-supervised text model, for generating augmented speech from emotionally aligned sentences. This strategy significantly enriches the long audio training data, thereby improving the naturalness of the synthesized speech.
  \item We model the silent pause between utterances during data augmentation, hence, the model is encouraged to learn natural inter-sentence pausing. 
  \item We utilize self-contrastive learning to extract a more precise representation of speaking style from audio speech. 
\end{enumerate}

Together, these innovations contribute to the successful training of a more expressive model for audiobook generation.

\section{Related Work}
\label{sec:related}

Following the significant advancements brought by Tacotron2 in the naturalness of machine-generated speech, efforts have been made to enhance the expressiveness of synthesized speech. For instance, Wang et al. ~\cite{ gsttaco} introduced the concept of global style tokens (GST), derived from expressive speech. The TTS model learns to combine these style tokens, forming a style embedding. This embedding is then used by the TTS model to generate expressive speech. Expanding on this idea, Stanton et al.~\cite{ stanton2018predicting} proposed a Text-Predicted Global Style Token (TP-GST) module, to predict the style embedding directly from the text. Various following researches~\cite{mello, kwon2019effective, bian2019multi} utilize GSTs on modeling speech's emotion. 



Raitio et al.~\cite{raitio2022improving} used a speech dataset consisting of short audio clips and long-form audio to train a TTS model. Their findings revealed that the synthesized speech with the long-form speaking style is favored even in short speech synthesis scenarios. Zhang et al.~\cite{zhang2023audiobook} upsampled BERT word embeddings~\cite{devlin2018bert} to match the phoneme-level embeddings and incorporated TP-GST into their proposed TTS model. They combined two consecutive sentences to create a long-context training dataset. Instead of using text to predict the style embedding, Lei et al.~\cite{lei2023context} fused the style embeddings of previous sentences and sentence embeddings of neighboring sentences to predict the ground truth style embedding of the target utterance. Xiao et al.~\cite{xiao2023contextspeech} incorporated BERT-based word embeddings and sentence embeddings, and token-level statistical features to a TTS model. By applying linearized self-attention, their model could vocalize up to 1500 input Chinese phone numbers. Nevertheless, If the input exceeds the limit of a TTS model, it would necessitate concatenating synthesized speech segments.

Recently, there has been increased interest in using self-supervised learning (SSL) features directly for speech synthesis. Fujita et al.~\cite{fujita2023zero} utilize the weighted sum of SSL's different layer outputs to obtain an embedding vector, finding it superior to x-vector speaker embedding and suitable for an extra input to duration prediction. Shi et al.~\cite{shi4341256wfsc} employ frames similarity and SSL features to encode Chinese characters, replacing the pronunciation dictionary lookup with a codebook. Lim~\cite{lim2021preliminary} and Siuzdak~\cite{siuzdak2022wavthruvec} replace the popular Mel-spectrogram with wav2vec2~\cite{baevski2020wav2vec} embeddings as an intermediate speech representation in TTS model training. In this paper, we propose using a fine-tuned text SSL for speech training data augmentation.

SimCLR\cite{chen2020simple} is an unsupervised learning framework that employs contrastive learning for training visual representations. This framework can be adapted to other domains by modifying the base encoder network and incorporating domain-specific data augmentation techniques, enabling it to learn more general and invariant features, particularly when faced with limited data. Several studies\cite{jiang2020speech, khorram2022contrastive} have successfully applied contrastive learning to speech data, resulting in enhanced speech representations for speech recognition. As far as we are aware, our work represents one of the initial attempts to apply this technique to speaking style representation extraction for expressive speech synthesis.

In this study, we introduce a straightforward yet effective data augmentation method for training expressive TTS models. We propose forming long-form audio from emotionally matching sentences rather than consecutive sentences to leverage the advantages of using long-form speech data in training a TTS model. Furthermore, recognizing the strength of pre-trained SSL models in various cases, instead of combining pre-trained BERT word embeddings with phone embeddings, we opt to employ a fine-tuned SSL model for classifying sentences and producing emotion-coherent augmented data. Furthermore, we consider inter-sentence pausing as an integral component of speech synthesis. Hence, our model can be used to deduce suitable pauses between paragraphs. 

\section{Proposed Method}
\label{sec:proposed}

We focus on optimizing the Global Style Tokens module, often used to boost Fastspeech2 and Tacotron2 expressiveness. We base our work on Tacotron2 instead of Fastspeech2 as Fastspeech2's design limits its expressive speech synthesis capabilities. Its parallel generation of speech features, as opposed to sequential, impedes capturing speech dependencies. The pre-decoupling of acoustic features necessitates a forced aligner, which is prone to misalignments in expressive speech. Tacotron2 takes a phoneme sequence as input and produces a Mel-spectrogram as output. To ensure robust and longer speech synthesis, we utilize the stepwise monotonic attention mechanism and apply a reduction factor of 2 to extend the synthesized speech length. Furthermore, we incorporate TP-GST \cite{tpgst} to extract expressive speaking styles from the Mel-spectrogram and predict those styles from text, allowing for the generation of expressive speech for an unseen text. To convert the Mel-spectrogram into a waveform, we utilize a pre-trained WaveGlow \cite{waveglow} vocoder. 


\subsection{Data Augmentation by Speech Concatenation}
Markarov et al. \cite{makarov2022simple} posit that training a TTS model with utterances comprising multiple coherent adjacent sentences results in improved prosody in the synthesized speech. In their study, they experimented with a proprietary 78-hour dataset, chunking the data into approximately 24-second segments, which equates to around three sentences. Generally, expressive voice training data is scarce. In this work, we propose a straightforward method to augment the dataset. First, we divide the training audio data into individual sentences. Next, we concatenate the utterances to generate long-form speech for training purposes. As for the text, we insert a space character between sentences and an additional special symbol \~{} to signify the end of the training text. We incorporate suitable pauses between utterances, based on a normal distribution derived from actual data.

\subsection{Sentence Emotion Recognition}

T5~\cite{Raffel2019ExploringTL}, also known as Text-to-Text Transfer Transformer, is a pre-trained encoder-decoder model using a mix of unsupervised and supervised tasks. It can be fine-tuned for various natural language processing tasks, including sentence emotion recognition. Instead of random audio concatenation, we use a text classifier to detect emotions in sentences and concatenate speeches with matching emotion labels. This is crucial as emotions in consecutive sentences can vary, especially in story character conversations. Utilizing utterances with varying emotions could negatively impact GST's capacity to accurately learn style tokens from expressive speeches. Moreover, concatenating audios with the same emotion offers additional data for global style tokens training and suitable speaking style extraction.


\subsection{Self-Supervised Contrastive Style Training}

In the context of GST-based speech synthesis models, the mel-spectrogram of an utterance is fed into a reference encoder to extract a query embedding for weighing style tokens to form a speaking style embedding. This embedding is subsequently conditioned by the TTS decoder to generate expressive speech. To augment the speech data, we introduce random masking of 500 ms segments. We refrain from applying pitch shifting or speech perturbation, as these techniques could significantly alter the speaking style of the original utterance. We create two views for each sample within the same training batch. We apply the SimCLR loss on the reference encoder of the GST module during training, thereby enabling the reference encoder to extract more consistent global style embedding from the different masked views of the reference audios.

\section{Experiments and results}
\label{sec:expt}

\subsection{Dataset}

We initially pre-trained our TTS model using the LJS corpus \cite{ljspeech17}, which comprises approximately 21 hours of speech from a native English female speaker. We trained for 300 epochs. Subsequently, we employed the LibriTTS dataset \cite{zen2019libritts}, filtered and processed following the same procedure as Mellotron \cite{mello}, to train the TTS model for 200 epochs, enabling the development of style tokens and TP-GST module. Finally, we fine-tuned the model with the target speaker – a female child storyteller from the 2017 Blizzard Machine Learning Challenge for 100 epochs.

The Blizzard Challenge 2017 dataset includes a female English speaker narrating 61 children's stories. We designated one story (Three Little Pigs) as the test set and used remaining stories for training. Excluding intro and outro speech, we had 5,155 sentences. The dataset contains manual labels for sentence start and end times. We calculated time differences between consecutive sentence pairs, excluding those with a difference over 1 second, considering them outliers. Fitting a normal distribution, we obtained a mean of 509ms and std of 223ms for inter-sentence pauses.




We segment the audio into individual sentences, into 2 consecutive sentence audio and 3 consecutive sentence audio, the audio length statistic is summarized in Table 1. The average length of a 1-sentence utterance is less than 3, indicating that a model trained solely on 1-sentence utterances may not perform adequately for longer text. However, if the data is divided into 3-sentence utterances, the quantity of training data is reduced by nearly half. We refrain from merging the raw audio from consecutive pages, as there is no information about the natural pauses between them.

\begin{table}
\centering
\label{table:3}
\begin{tabular}{c|cccl} 
\hline
                 & \multicolumn{4}{c}{\textbf{Audio Length in seconds}}  \\ 
\hline\hline
\textbf{Audio} &  Count & Mean  & Min.   &     Max.              \\ 
\hline
1-sentence utterance &  5150  & 2.64 & 0.24 &      15.56             \\ 
\hline
2-sentence utterance &  3587            & 5.60 & 1.05 &      17.64            \\ 
\hline
3-sentence utterance &   2409      &  8.47 & 2.47 &      20.42             
\\
\hline
\end{tabular}
\caption{Statistic of the segmented speech data.}
\label{table:wer}
\end{table}

Furthermore, we use a SSL model fine-tuned emotion recognition model \footnote{https://huggingface.co/mrm8488/t5-base-finetuned-emotion} to predict the label of each sentence. It was trained with a large dataset \cite{saravia-etal-2018-carer}. This model achieved 93\% accuracy in its test set. We selected 200 random sentences from the LJSpeech corpus, which is regarded as a non-fiction dataset, and computed the average scores of the predicted emotion labels using this classifier. The average score was found to be 0.65. We applied a higher threshold of 0.7, categorizing scores below this value as neutral. The distribution of the predicted labels for the Blizzard Challenge 2017, based on the aforementioned threshold, is depicted in Table 2. Table 3 provides an example for each emotionally-labelled sentence. Furthermore, based on our labels, 41\% of them are non-neutral, and we found that a mere 8\% of consecutive sentences have the same non-neutral emotion labels. Directly segmenting longer utterances from the dataset might not seem like the most effective strategy for a GST-based speech synthesis model.


\begin{table}
\centering
\label{table:3}
\begin{tabular}{c|cccl} 
\hline
Emotion & Neutral &  Joy & Fear  & Anger         \\ 
\hline
Percentage & 58.75 &  16.90 & 8.62 & 10.92           \\ 
\hline\hline
Emotion &     Sadness   &     Love   &     Surprise   &           \\ 
\hline
Percentage &  3.05 &     1.16 &     0.61    &       \\     
\hline
\end{tabular}
\caption{Distribution of the emotion classification of the training sentences.}
\label{table:emd}
\end{table}


\begin{table}[tbp]
\centering
\begin{tabular}{c|p{60mm}} 
\hline
Emotion & Samples                 \\ 
\hline\hline
Joy               &    But there was a delicious smell coming from inside.       \\
\hline
Fear               &      A lion is watching the elephant babies.     \\
\hline
Anger               &     "Who's been eating my porridge?" he said, in his great, gruff voice.       \\
\hline
Sadness               &    Alone again, Macbeth thought of his best friend Banquo, who would soon be dead.       \\
\hline
Love               &     Elephants like to live in hot places.       \\
\hline
Surprise               &   Helena was shocked.        \\
\hline
\end{tabular}
\caption{Examples of sentences labeled by the model.}
\label{table:es}
\end{table}




\subsection{Model Training}

We modified the Mellotron\cite{mello} official codebase to implement our idea. We applied various training datasets with real and in-memory augmented data and incorporated self-contrastive training. 




\noindent \textbf{M1}: Trained using only 1-sentence utterances.

\noindent \textbf{M2}: Trained using both 1-sentence utterances and 2-sentence utterances of the original consecutive sentences.  



\noindent \textbf{M3}: Trained using 1-sentence utterances and augmented 2-sentence utterances created by concatenating 1-sentence utterances with matched emotion labels.

\noindent \textbf{M4}: Trained using 1-sentence utterances and augmented 2-sentence utterances created by concatenating 1-sentence utterances with matched emotion labels, and applied the self-contrastive training as discussed in Section 3.3. We incorporate the SimCLR loss into the total training loss of the TTS model, applying a scaling factor of 0.1.

Furthermore, for models M2, M3, and M4, the concatenated two sentences serve as a single input for the text encoder. Afterward, we separated the text embedding based on the two sentences and employed the TPGST to predict their speaking style embeddings. In this case, each sentence has its style embedding. By freezing these embeddings and passing them to the TTS decoder, we were able to simulate the inference process and optimize the TTS decoder based on the predicted TPGST. In half of the cases, we utilized the GST embeddings obtained directly from the combined emotion-coherent speech data. This additional emotional data aids in optimizing the GST module more effectively.


\noindent Some audio samples are available on this webpage\footnote{https://raymond00000.github.io/storyttsdemo.html}.



\subsection{Objective Evaluation}



In Table 4, we present some objective evaluations. Firstly, we evaluate the speaking style embedding, anticipating that its representation will be enhanced by our proposed data augmentation strategy and self-supervised contrastive learning. The TP-GST module predicts the speaking style embedding from text to match that extracted from audio. We found that the model (M3, M4) trained with our augmented data has a lower prediction error on the test set data. 

We selected a female speaker from a speech dataset, the Emotional Speech Dataset (ESD)~\cite{zhou2021emotional}, which was unseen by our model. The speaker recorded 350 utterances for each of the five emotional states (neutral, happy, angry, sad, and surprised). These audio samples were passed through the trained GST extractor to obtain speaking style embeddings. Subsequently, we employed a Support Vector Machine classifier to predict the emotion label. With a 25\% test set random partition, this process was repeated 50 times, and the average accuracy on the test set was calculated. Our proposed models (M3, M4) exhibited higher accuracy, with M4 indicating that self-contrastive learning could further reduce the L1 loss while slightly enhancing emotion classification. These findings suggest that utilizing self-contrastive learning can allow the GST module to extract a better speech representation for emotional expressiveness.


Secondly, we evaluate the model's ability to effectively learn inter-sentence pausing. To achieve this, we incorporate silent pauses between utterances in two-sentence samples, challenging the TTS model to learn how to insert appropriate pauses between multiple lines of speech. We utilize the M1 model, which is trained with one-sentence utterances only, and the M3 model, which is trained with inserted inter-sentence pauses drawn from a normal distribution. Both models are used to generate two-sentence utterances in the test set. We utilized the Montreal Forced Aligner~\cite{mcauliffe17_interspeech}, a forced alignment tool, along with a pre-trained English acoustic model~\footnote{English MFA acoustic model v2.0.0}, to align speech and text, enabling us to determine the inter-sentence pause length. 

We compared the distribution of M1 and M3 against the ground truth of inter-sentence pause data using the two-sample Kolmogorov-Smirnov (K-S) test. The results for M3 show a K-S statistic of 0.247 and a p-value of 0.630, while M1 exhibits a K-S statistic of 0.490 and a p-value of 0.0271. Since the K-S statistic is smaller for the M3 model, it can be concluded that M3 generates more natural inter-sentence pauses than the M1 model. This finding suggests that we could incorporate natural-sounding pausing information into the TTS model via a data augmentation training process. 

In particular, since there is an upper limit on the length of synthesized audio from a TTS model, audio concatenation is inevitable. Using an autoregressive model, the system can be initialized with an input text and audio pair, followed by decoding subsequent sentences. Our TTS model could determine the optimal pause duration between consecutive audio segments, thus enhancing the overall perceptive naturalness of the synthesized speech for audiobook creation.




\begin{table}[tbp]
\centering
\begin{tabular}{c|cccl} 
\hline
Model & M1  & M2   &     M3   &     M4                  \\ 
\hline\hline
L1 loss on TPGST               & 0.212 & 0.155   & 0.119  & 0.075          \\ 
\hline
Emotion classification & 71.5\% & 70.7\%   & 75.1\%  & 75.3\%        \\ 
\hline
\end{tabular}
\caption{L1 loss on TPGST, Extracted speaking style embedding for emotion classification accuracy of different models on the test data.}
\label{table:ss}
\end{table}



\subsection{Subjective Evaluation}

We conducted a crowdsourcing mean opinion score (MOS) evaluation, following the same settings as the rating task in Clark et al.~\cite{clark2019evaluating}. We recruited 8 native speakers from the United Kingdom via Prolific\footnote{https://www.prolific.com/} to participate in the study. Participants were asked to assess the naturalness of the speech, i.e., whether it sounded as if spoken by a human, as well as the appropriateness of the speech, in terms of its suitability for the given context. The evaluation scale ranged from 1 to 5, with 0.5 increments. The results are presented in Table 5. Our sample model M4 has outperformed the baseline model M2 by achieving a higher point estimate. As we increase the number of testers, we anticipate a reduction in the margin of error. This trend underscores the efficacy of our proposed methodology. While additional training data could improve the results, we limited ourselves to a well-curated, highly challenging academic dataset, promoting comparisons among researchers.

Moreover, we observed that variations in the F0 range impact speech quality, likely due to the absence of higher F0 values in the vocoder training data. This could be attributed to the lower naturalness of our synthesized samples. We suggest that fine-tuning the vocoder could potentially further enhance speech quality.



\begin{table}[tbp]
\centering
\begin{tabular}{l|ll} 
\hline
Audio        & Naturalness & Appropriateness   \\ 
\hline
\hline
M2 & 3.19±0.56           & 3.36±0.45           \\
M4  & 3.25±0.56             & 3.42±0.52           \\
\hline
\end{tabular}
\caption{MOS results at 95\% confidence level.}
\label{table:mos}
\end{table}




\section{Conclusion}
\label{sec:conc}

Our study makes two main contributions to the field of kids's story speech synthesis research. First, we have developed an emotionally matching data augmentation strategy that leverages a fine-tuned text-based emotion predictor from a self-supervised learning model, resulting in the effective expansion of audio training data with more emotionally expressive speeches. Second, our findings suggest that self-contrastive learning can assist in training a more refined speaking style embedding, which conditions the decoder of speech synthesis for expressive speech.

By employing these strategies, we have successfully trained an efficient expressive speech model, as evidenced by the reduced speaking style embedding prediction loss on test data and a higher emotion speech prediction accuracy for an unseen speaker compared to the baseline model. Subjective evaluations further corroborate the enhanced naturalness and speaking style appropriateness of synthesized speech generated by our model compared to the baseline model. Additionally, incorporating silent pauses between two-sentence samples fosters more natural inter-sentence pauses within the TTS model.

For future research, employing a state-of-the-art text-based SSL model such as GPT, renowned for its exceptional performance in a wide range of NLP tasks, has the potential to enhance emotion prediction accuracy and produce improved emotion-matching augmented data for training increasingly sophisticated expressive speech models with limited data. Furthermore, we plan to assess the efficacy of our proposed model across various languages and scenarios. 




\section{ACKNOWLEDGMENTS}
\label{sec:ack}

This work was supported by a grant from the Innovation and Technology Fund of the Hong Kong SAR, China (Project No: ITP/046/23LP, ITT/008/23LP).

\clearpage
\bibliographystyle{IEEEbib}
\bibliography{strings,refs}

\end{document}